\documentstyle[12pt]{article}

\newcommand{\be}{\begin{equation}}
\newcommand{\ee}{\end{equation}}
\newcommand{\bea}{\begin{eqnarray}}
\newcommand{\eea}{\end{eqnarray}}
\newcommand{\vs}[1]{\vspace{#1 mm}}

\def\bbox{{\,\lower0.9pt\vbox{\hrule \hbox{\vrule height 0.2 cm
\hskip 0.2 cm \vrule height 0.2 cm}\hrule}\,}}
\newcommand{\dsl}{\pa \kern-0.5em /}

\newcommand{\pa}{\partial}

\newcommand{\nn}{\nonumber\\}

\font\mybb=msbm10 at 12pt
\def\bb#1{\hbox{\mybb#1}}

\def\bR {\bb{R}}
\def\bE {\bb{E}}

\begin{document}

\topmargin 0pt
\oddsidemargin 5mm

\renewcommand{\thefootnote}{\fnsymbol{footnote}}
\begin{titlepage}

\setcounter{page}{0}
\begin{flushright}
DAMTP-R/98/05 \\
hep-th/9801165
\end{flushright}

\vs{5}
\begin{center}
{\Large Gauged supergravity vacua from intersecting branes}
\vs{10}

{\large
P.M. Cowdall
} \\
\vs{5}
{\em DAMTP, University of Cambridge, \\
Silver Street, Cambridge, U.K.} \\
\vs{7}
and \\
\vs{7}
{\large
P. K. Townsend\footnote{On
leave from DAMTP,  University of Cambridge, U.K.} } \\
\vs{5}
{\em Departament ECM, Facultat de F{\'\i}sica, \\
Universitat de Barcelona and Institut de F{\'\i}sica d'Altes Energies, \\
Diagonal 647, E-08028 Barcelona, Spain }\\ 
\end{center}
\vs{10}
\centerline{{\bf{Abstract}}}

Domain wall and electrovac solutions of gauged N=4 D=4 supergravity, with gauge
group $SU(2)$ or $SU(2)\times SU(2)$, are interpreted as supersymmetric
Kaluza-Klein vacua of N=1 D=10 supergravity. These vacua are shown to be the
near-horizon geometries of certain intersecting brane solutions.

\end{titlepage}
\newpage
\renewcommand{\thefootnote}{\arabic{footnote}}
\setcounter{footnote}{0} 

\section{Introduction}

Gauged supergravity theories are those for which a subgroup of the R-symmetry
group (alias the automorphism group of the supersymmetry algebra) is gauged by
vector potentials in the graviton supermultiplet. The gauging is invariably
accompanied by a cosmological constant of order $g^2$, where $g$ is the gauge
coupling constant, and in the simplest cases (the `adS-supergravities') 
there is a maximally supersymmetric anti-de Sitter (adS) vacuum, maximal in the
sense that the adS vacuum of the gauged supergravity theory preserves the same
total number of supersymmetries as the Minkowski vacuum of the ungauged theory.  
A classic example is gauged N=8 D=4 supergravity, which can be obtained by
gauging an $SO(8)$ subgroup of the $SU(8)$ R-symmetry group. This theory has an
adS vacuum with the N=8 adS supergroup $OSp(8|4;\bR)$ as its isometry supergroup.
It can also be obtained by $S^7$ compactification of D=11 supergravity, in which
case the $SO(8)$ gauge group has a Kaluza-Klein (KK) origin as the isometry
group of $S^7$. Another example is the gauged  D=5 maximal supergravity for
which the gauge group is an $SU(4)$ subgroup of the $Sp(4)$ (alias $USp(8)$)
R-symmetry group. This too has an adS vacuum and can be obtained from an $S^5$
compactification of IIB supergravity.  Yet another example is the gauged maximal
D=7 supergravity, for which the gauge group is the full $Sp(2)\cong Spin(5)$
R-symmetry group. This theory has the curious feature that the $g\rightarrow 0$
limit is singular, so it cannot be found by the usual `Noether' procedure, in
which the ungauged theory is taken as the starting point; it was actually found
from an $S^4$ compactification of D=11 supergravity. We refer
to \cite{duffpope,pkta} for reviews and references to  work of this period.
 
It was shown in \cite{gibtown} that all the above mentioned KK vacua associated
with gauged maximal supergravities arise as near-horizon geometries of 1/2
supersymmetric p-brane solutions of D=10 or D=11 supergravity theories.
The $adS_4\times S^7$ and $adS_7\times S^4$ vacuua of D=11 supergravity are the
near-horizon geometries of the extreme membrane and fivebrane solutions,
respectively, while the $adS_5\times S^5$ vacuum of IIB supergravity is the
near-horizon geometry of the threebrane solution. In other words, these p-brane
solutions interpolate between maximally supersymmetric vacua of the
respective supergravity theories. More recently it has been shown that some
intersecting brane solutions have a similar property. For example, the extreme
black hole and black string solutions of D=5 minimal supergravity \cite{ght}
(reducing to $a=1/\sqrt{3}$ dilaton black holes in D=4) interpolate between
either $adS_2\times S^3$ or $adS_3\times S^2$, respectively \cite{ferrara,
behrndt,skenderis}, but these are the reduction of D=11 supergravity
solutions that can be interpreted \cite{paptown} as, respectively, three
intersecting M2-branes or three intersecting M5-branes. Other examples have
been given in \cite{kallosh,skenderis}, and it was shown more generally in
\cite{skenderis} that intersecting M-branes interpolate between the D=11
Minkowski vacuum and a spacetime of the form $adS_k\times \bE^l\times S^m$ (with
$k+l+m=11$). 

In this paper we shall explore similar issues in the context of N=1 D=10
supergravity. Some observations concerning this case have been made previously
in the context of black hole entropy \cite{hyun,skenderis,sfetsos}, and the
topic has been revitalized by recent conjectures relating near-horizon
geometries to large rank limits of supersymmetric gauge theories \cite{malda},
but our principal concern is to explore some connections to the N=4 D=4 gauged
supergravity theory of Freedman and Schwarz \cite{FS}, which we call the `FS
theory'. The gauge group is $SU(2)\times SU(2)$ with gauge coupling constants
$e_A$ and $e_B$, unless $e_B=0$ in which case the gauge group is $SU(2)\times
U(1)^3$ with $e_A\equiv g$ being the $SU(2)$ gauge coupling constant; we shall
call the latter theory the `half-gauged' FS model. There is another N=4 D=4
gauged supergravity \cite{gates}, usually called the `$SO(4)$ theory', which has
the same field content and gauge group but different interactions, except for the
`half-gauged' case which coincides with the half-gauged FS model. Only the FS
model, gauged or half-gauged, will be of relevance here. Its distinguishing
feature is that the single scalar field $\sigma$ (the D=4 dilaton) has a
potential 
\be
\label{eq:onea}
V = 2(e_A^2+e_B^2)e^\sigma
\ee
so that there is no Minkowski vacuum. This feature is also shared by gauged D=7
minimal supergravity \cite{vanN} (with vanishing topological mass term
\cite{luca}). This is no coincidence as the `half-gauged' FS model is the
dimensional reduction on $T^3$ of the D=7 theory \cite{cowdall}.

It is natural to suppose that gauged D=7 supergravity is an $S^3$
compactification of D=10 N=1 supergravity since the $SU(2)$ gauge group would
then acquire a KK origin as one factor of the $SU(2)\times SU(2)$ isometry
group of $S^3$ (the gauge fields of the other $SU(2)$ factor would have to
belong to three vector multiplets which could likely be consistently truncated).
If so, the FS model would then have a natural KK interpretation as an $S^3\times
S^3$ compactification of D=10 N=1 supergravity (followed by a truncation of an
$SU(2)\times SU(2)$ Yang-Mills multiplet). The `half-gauged' FS model would
then acquire a similar interpretaton as an $S^3\times T^3$ compactification.
These suppositions are in fact correct, although it was a long time before this
was appreciated \cite{bachas,cham}. One reason for the delay is that the 
first $S^3$ compactification of D=10 N=1 supergravity to be found \cite{dtv} 
is such that $\Phi\equiv e^\phi$ (where $\phi$ is the D=10 dilaton) is not
everywhere positive on $S^3$. The analogous $S^3\times S^3$ compactification
suffers from the same problem, and requiring positivity of $\Phi$ led, in
1983, to a `no-go' theorem that apparently precluded the existence of a
physically acceptable $S^3\times S^3$ compactification to D=4 \cite{gfw}.

There were no further attempts to provide a KK origin for the FS model until
1990, when the FS model was identified as part of the effective D=4 field
theory for the heterotic string theory in an $S^3\times S^3$ vacuum
\cite{bachas}. The `no-go' theorem is circumvented by the fact that the D=4
dilaton is not presumed to be constant. In a subsequent independent development,
it was discovered \cite{gibtown} that the (non-singular) fivebrane solution of
D=10 supergravity \cite{duff,chs} interpolates, in the string metric, between
the Minkowski vacuum and an $S^3$ compactification to D=7 Minkowski spacetime,
and it was noted that an $S^3\times S^3$ compactification to D=4 Minkowski
spacetime is also possible. Again, neither the D=7 nor the D=4 dilaton is
constant in these compactifications but, rather, linear in one of the Minkowski
coordinates. We shall show here that the D=7 linear dilaton vacuum is actually
the 1/2 supersymmetric domain wall solution of gauged D=7 supergravity found in
\cite{pope}. This reduces in D=4 to a 1/2 supersymmetric domain wall solution of
the half-gauged FS model, which is also a 1/2 supersymmetric solution of the
full $SU(2)\times SU(2)$ FS model \cite{cowdall}. There is therefore no obstacle
to the identification of gauged D=7 supergravity as a consistent truncation of
$S^3$ compactified D=10 N=1 supergravity, and of the $SU(2)\times SU(2)$ FS
model as a consistent truncation of $S^3\times S^3$ compactified D=10 N=1
supergravity (the truncations merely removing inessential matter multiplets).
The latter identification was made and verified in \cite{bachas,cham}. The 
former identification (conjectured in \cite{skenderis}) then follows from the
results in \cite{cowdall}. 

Here we consider further these $S^3$ and $S^3\times S^3$ compactifications of
D=10 supergravity (N=1 is henceforth assumed). We take the D=10 action to be 
\be
\label{eq:oneb}
S = \int d^{10} x\; \sqrt{-g}e^{-2\phi}[R + 4(\partial\phi)^2 -{1\over12}F^2]
\ee
where $\phi$ is the dilaton and $F$ is the 3-form field strength. The metric is
thus the string-frame metric. We shall show that the $S^3\times S^3$
compactification of this theory is the near horizon geometry of a solution
representing the intersection of two fivebranes on a line. The latter
solution \cite{khuri}, which preserves 1/4 supersymmetry, therefore interpolates
between the fully supersymmetric Minkowski vacuum and the 1/2 supersymmetric 
$S^3\times S^3$ compactification to the domain wall of the FS model. 

The domain wall is not the only 1/2 supersymmetric solution of the FS theory.
There is a 1/2 supersymmetric electrovac solution of the half-gauged model
\cite{gibfreed}. This was shown in \cite{cowdall} to descend from an
analogous `electrovac' solution of gauged D=7 supergravity, for which the
metric is actually just $(adS)_3\times \bE^4$, corresponding to an
$adS_3\times \bE^4 \times S^3$ solution of D=10 supergravity. Unlike the domain
wall, the `electrovac' solution has a constant dilaton. Here we shall
show that it is the near-horizon geometry of a 1/4 supersymmetric intersecting
brane solution of D=10 supergravity in which a string lies inside a 
fivebrane\footnote{This fact was also noted in \cite{aat,skenderis}, but 
without the connection to the FS model. A related observation was made in
\cite{hyun}.}. Alternatively, by replacing  $\bE^4$ by $T^4$ in the
D=10 `electrovac' solution, we can consider it as an $adS_3\times S^3$ solution
of the dimensionally reduced D=6 supergravity. This solution is the near horizon
geometry of the self-dual D=6 string of \cite{dufflu}, which is the reduction to
D=6 of the D=10 intersecting brane solution. Considered in the context of {\sl
minimal} D=6 supergravity, the self-dual D=6 string is similar to the M2-brane,
M5-brane and D3-brane in that it is a 1/2 supersymmetric solution that
interpolates between the fully supersymmetric $Mink_6$ and $adS_3\times S^3$
vacua of this theory\footnote{See \cite{toine} for recent
related observations.}. Thus, surprisingly, the Gibbons/Freedman electrovac of
gauged D=4 supergravity is directly related to the D=6 self-dual string.

The D=10 string-in-fivebrane solution can be generalized to a string in the
intersection of two fivebranes (which we choose to be orthogonal); this 1/4
supersymmetric solution can be found by an application of a `generalized
harmonic function rule' of \cite{aat,tsey,GGPT}. By an appropriate choice of the
harmonic functions one can arrange for the dilaton to be constant and for the
metric to interpolate between the Minkowski vacuum and an $S^3\times S^3$
compactification to $adS_3 \times \bE^1$. This establishes the existence of a
supersymmetric $adS_3\times \bE^1$ vacuum of the $SU(2)\times SU(2)$ FS model
with at least 1/4 supersymmetry. It is presumably the 1/4 supersymmetric
`axionic' solution recently found in \cite{singh}. 

\section{Domain walls from intersecting fivebranes}

The 1/4 supersymmetric solution of D=10 supergravity (in string frame)
representing two orthogonal fivebranes intersecting on a line is
\begin{eqnarray}
\label{eq:twoa}
ds^2 &=& ds^2(\bE^{(1,1)}) + H dx\cdot dx + H' dx'\cdot dx' \nn
e^{2\phi}&=& HH'\nn 
F &=& \star dH + \star' dH'
\end{eqnarray}
where $\bE^{(1,1)}$ indicates a (1+1)-dimensional Minkowski space, $H$ and $H'$
are harmonic functions on their respective 4-dimensional Euclidean spaces with
metrics $dx\cdot dx$ and $dx'\cdot dx'$, and $\star$ and $\star'$ are the Hodge
duals on these two spaces. We choose the harmonic functions to be
\be
\label{eq:twob}
H = 1 + {1\over r^2} \qquad H' = 1+ {1\over r'^2}
\ee
where $r = |x|$ and $r' = |x'|$ are the distances from the origins of the
two 4-dimensional Euclidean spaces. 

Close to the first fivebrane, but far from the second one, we have $H\sim
1/r^2$ and $H' \sim 1$. In this case, the asymptotic metric is \cite{gibtown}
\begin{eqnarray}
\label{eq:twoc}
ds^2 &\sim& ds^2(\bE^{(1,1)}) + dx'\cdot dx' +  {dr^2\over r^2} + ds^2(S^3)\nn
&=& ds^2(\bE^{(1,6)}) + ds^2(S^3)
\end{eqnarray}
while the dilaton is $\phi \sim \rho$, where $\rho =-\log r$. From the 
discussion in the introduction we now deduce that this $S^3$ compactification of
D=10 supergravity implies the existence of a solution of D=7 gauged supergravity
with dilaton $\phi= \rho$, Minkowski 7-metric and vanishing D=7 gauge fields.
Using the relation $ds^2_E = e^{-4\phi/5} ds^2$ between the string-frame
7-metric  and the Einstein-frame 7-metric we find that this solution
has Einstein-frame 7-metric
\be
\label{eq:twoe} 
ds^2_E = e^{-{4\over 5}\rho} [ds^2(\bE^{(1,5)}) + d\rho^2]
\ee
Defining $y = (1/2)e^{2\rho}$ we find the solution
\begin{eqnarray}
\label{eq:twof}
ds^2_E &=& H^{-{2\over5}} ds^2(\bE^{(1,5)}) + H^{-{12\over5}} dy^2 \nn
e^{2\phi} &=& H
\end{eqnarray}
with $H=2y$. 

To compare with the solutions of gauged D=7 supergravity
we need to write the D=10 dilaton $\phi$ in terms of the D=7 dilaton 
$\phi_{(7)}$. Later we shall need to do the same for the D=4 dilaton 
$\phi_{(4)}\equiv \sigma$. We therefore pause here to deduce the relation
between $\phi$ and the D-dimensional dilaton $\phi_{(D)}$. Since the
10-metrics we consider are direct products of D-dimensional metrics
with spheres, or products of spheres, of constant radius, the D=10
dilaton $\phi$ remains the only scalar field in the lower dimension,
so that the effective D-dimensional lagrangian is still of the form 
\be
\label{eq:twofa}
L = \sqrt{-g}e^{-2\phi} [R+4(\nabla\phi)^2 + \dots ]\, .
\ee
This is equivalent to 
\be
\label{eq:twofb}
L_E = \sqrt{-g_E} [R - {1\over 2} (\nabla\phi_{(D)})^2 + \dots ]_E \, ,
\ee
where  
\be
\label{eq:twofc}
\phi_{(D)} = {2\sqrt{2}\over \sqrt{D-2}}\, \phi\, ,
\ee
and the subscript `$E$' indicates a D-dimensional Einstein-frame metric.
Thus, for D=7 we have
\be
\label{eq:twofd}
\phi_{(7)} = {2\sqrt{2}\over \sqrt{5}}\, \phi.
\ee
Using this relation one sees that the solution (\ref{eq:twof}) is the 1/2
supersymmetric domain wall solution of \cite{pope} (for which, in general, 
$H$ is piecewise linear in $y$). 

We now turn to the asymptotic metric in the region near both fivebranes. In
this case $H\sim 1/r^2$ and $H'\sim 1/r'^2$. Setting 
\be
\label{eq:twofe}
\rho = -{1\over\sqrt{2}}\log (rr') \qquad 
\lambda = {1\over\sqrt{2}}\log (r/r')\, ,
\ee 
we then find 
\begin{eqnarray}
\label{eq:twog}
ds^2 &\sim& ds^2(\bE^{(1,1)}) + d\rho^2 + d\lambda^2 + ds^2(S^3\times S^3) \nn
\phi &\sim& \sqrt{2}\rho
\end{eqnarray}
while the 3-form field strength is now the sum of the volume forms of the two
$S^3$ factors. This result implies the existence of a supersymmetric 
solution to
the D=4 $SU(2)\times SU(2)$ FS model with vanishing gauge fields.
Passing to the  (D=4) Einstein frame, for which $\phi_{(4)}\equiv \sigma =
2\phi$, and defining $y=(1/2\sqrt{2})e^{2\sqrt{2}\rho}$, we find that
\begin{eqnarray}
\label{eq:twoh}
ds^2_E &=& H^{-1} ds^2(\bE^{(1,2)}) + H^{-3} dy^2\nn
e^{-\sigma} &=& H
\end{eqnarray}
with $H=\sqrt{2}y$. This is just the 1/2 supersymmetric domain
wall solution of \cite{cowdall}, shown there to be the dimensional reduction of
the D=7 domain wall solution. 

Thus, the intersecting fivebrane solution interpolates between the D=10
Minkowski vacuum and a 1/2 supersymmetric domain wall solution of either D=7
gauged supergravity or $SU(2)\times SU(2)$ D=4 gauged supergravity, according to
whether we are close to just one of the fivebranes or both of them. 

\section{Electrovacs from string-in-fivebrane}

The 1/4 supersymmetric string-in-fivebrane solution of D=10 supergravity is
\begin{eqnarray}
\label{eq:threea}
ds^2 &=& H_1^{-1}ds^2(\bE^{(1,1)}) + H_5 dx\cdot dx + ds^2(\bE^4) \nn
e^{2\phi} &=& H_1^{-1}H_5 \nn
F &=&  vol(\bE^{(1,1)})\wedge dH_1^{-1} + \star dH_5
\end{eqnarray}
where $H_1$ and $H_5$ are both harmonic functions on the 4-space with Euclidean
4-metric $dx\cdot dx$, and $\star$ is the Hodge dual on this space.  We shall
choose 
\be
\label{eq:threeb}
H_1 = H_5 = H(x)\, .
\ee
This choice has the property that the dilaton is constant, in
fact zero. The metric and 3-form field strength are now
\begin{eqnarray}
\label{eq:threec}
ds^2 &=& H^{-1} ds^2(\bE^{(1,1)}) + H dx\cdot dx + ds^2(\bE^4)\nn
F    &=&  vol(\bE^{(1,1)})\wedge dH^{-1} + \star dH
\end{eqnarray}
Ignoring the $\bE^4$ factor (which may be replaced by $T^4$), this
field configuration is automatically a 1/4 supersymmetric solution of the (1,1)
D=6 supergravity theory obtained by $T^4$ compactification of D=10 supergravity. 
Because the dilaton vanishes it is also a 1/2 supersymmetric solution of the
minimal (1,0) D=6 supergravity. In fact, it is just the self-dual string
solution of \cite{dufflu} (for which the singularities of $H$ were shown in
\cite{ght} to be horizons of the geodesically complete maximal analytic
extension). It then follows from the analysis below that the self-dual string
solution interpolates between the maximally supersymmetric Minkowski and
$adS_3\times S^3$ vacua of (1,0) D=6 supergravity. In this respect the D=6
self-dual string is smilar to the M2-brane, M5-brane and D3-brane. 

We now return to (\ref{eq:threec}) and choose
\be
\label{eq:threed}
H(x) = 1 + {1\over r^2}
\ee
where $r=|x|$. Near the origin the asymptotic metric is
\be
\label{eq:threee}
ds^2 \sim r^2 ds^2(\bE^{(1,1)}) + {dr^2\over r^2} + ds^2(S^3) + ds^2(\bE^4)
\ee
which is $adS_3 \times S^3 \times \bE^4$. The 3-form field strength $F$ is
asymptotic to the sum of the volume forms on the $S^3$ and $adS_3$ factors. 
By ignoring the $\bE^4$ factor, we deduce the result just claimed above for the
D=6 self-dual string. Instead, we may `ignore' the $S^3$ factor, i.e. we may
interpret the asymptotic solution just found as a new $S^3$ compactification of
D=10 supergravity preserving at least 1/4 supersymmetry. This implies the
existence of an $adS_3\times \bE^4$ vacuum of gauged D=7 supergravity,
again preserving at least 1/4 supersymmetry. It actually preserves 1/2
supersymmetry, so supersymmetry is partially restored near the horizon.
This follows from the fact that the `new' $adS_3\times \bE^4$ vacuum of gauged
D=7 supergravity is actually the D=7 `electrovac' found in \cite{cowdall}. The
D=7 `electrovac' metric is essentially of the form\footnote{Here we set
$g=\sqrt{2}$, set $\phi=0$, and choose horospherical coordinates.} 
\be
\label{eq:threef}
ds^2 = -e^{2\rho} dt^2 + d\rho^2 + (dy + e^\rho dt)^2 + ds^2(\bE^4)
\ee
The 3-form field strength of D=7 gauged supergravity is dual to a 4-form
field strength which, in this solution, is proportional to the
volume form on $\bE^4$. The $SU(2)$ gauge fields are zero. If we dimensionally
reduce on $y$ and two of the cartesian coordinates of $\bE^4$ then we recover
the 1/2 supersymmetric Gibbons-Freedman electrovac of the half-gauged FS model
(with $adS_2\times \bE^2$ 4-metric), hence the name given to the D=7 solution in
\cite{cowdall}. However, the 3-metric obtained by ignoring the $\bE^4$ factor is
just $adS_3$, as we have verified by a computation of the Ricci tensor. In
fact, the D=7 `electrovac' is equivalent to the compactification to $adS_3$
found in \cite{koreans}.

\section{`Axiovac' from string-in-two-fivebranes}
 
We now turn our attention to the solution representing a string in the
common linear-intersection of two fivebranes. The solution is 
\begin{eqnarray}
\label{eq:foura}
ds^2 &=& H_1^{-1}ds^2(\bE^{(1,1)}) + H_5 dx\cdot dx + H_5' dx'\cdot dx' \nn
e^{2\phi} &=& H_1^{-1}H_5 H_5'\nn
F &=&  vol(\bE^{(1,1)})\wedge dH_1^{-1} + \star dH_5 + \star' dH'_5
\end{eqnarray}
where $H_5$ is a harmonic function on the Euclidean 4-space with the $x$
coordinates, $H'_5$ is a harmonic function on the Euclidean 4-space with the
$x'$ coordinates, and the function $H_1$ satisfies \cite{tsey,GGPT}
\be
\label{eq:fourb}
\big[ (H_5)^{-1}\nabla^2 + (H_5')^{-1}\nabla'^2 \big]H_1 =0\, .
\ee
It was noted in \cite{GGPT} that this can be solved by additive separation of
variables. Of more relevance here is the fact that it can also be solved by
multiplicative separation of variables. Specifically, it is solved by
$H_1 = ff'$ where $f$ is harmonic in $x$ and $f'$ is harmonic in $x'$. In
particular, we may choose
\be
\label{eq:fourc}
H_1 = H_5H_5'\, .
\ee
This choice has the property that the dilaton is again zero. The other fields
are
\begin{eqnarray}
\label{eq:fourd}
ds^2 &=& (H_5H_5')^{-1} ds^2(\bE^{(1,1)}) + H_5 dx\cdot dx + 
H_5' dx'\cdot dx'\nn
F &=&  vol(\bE^{(1,1)})\wedge d(H_5H_5')^{-1} + \star dH_5 + \star' dH'_5
\end{eqnarray}
We now choose
\be
\label{eq:foure}
H_5 = 1 + {1\over r^2} \qquad H_5' = 1+ {1\over r'^2}\, .
\ee
Far away from one fivebrane, but close to the other one, we recover the
previous $adS_3 \times S^3 \times \bE^4$ solution. Near both fivebranes we 
use the coordinates (\ref{eq:twofa}) to write the asymptotic metric as
\be
\label{eq:fourf}
ds^2 = e^{-2\sqrt{2}\rho} ds^2(\bE^{(1,1)}) + d\rho^2 + d\lambda^2
+ ds^2(S^3\times S^3)
\ee
We recognize this as $adS_3 \times S^3\times S^3\times \bE^1$. The square of
the radius of curvature of the $adS_3$ factor is now half as large as before,
as required by the presence of two $S^3$ factors (given constant dilaton). The
original intersecting brane solution of D=10 supergravity preserves 1/4
supersymmetry, so the asymptotic solution near the fivebranes must also preserve
at least this fraction. It therefore corresponds to a solution of the D=4 FS
model that preserves at least 1/4 supersymmetry and has metric
$adS_3\times \bE^1$. This is presumably the 1/4 supersymmetric `axionic' 
vacuum solution, or `axiovac', of \cite{singh}.  
 
\section{Discussion}

We have shown that various supersymmetric vacua of the N=4 D=4 gauged
supergravity model of Freedman and Schwarz can be reinterpreted as
compactifications of D=10 N=1 supergravity, and that these compactifications
are the near-horizon geometries of various intersecting brane solutions.
The FS vacua that we can interpret in this way include the domain wall, the
$SU(2)\times U(1)^3$ electrovac, and the $adS_3\times \bE^1$ `axiovac'. 

There are other supersymmetric solutions for which we have not yet 
found a similar interpretation. An example which we believe should have such
an interpretation is the 1/4 supersymmetric electrovac of the $SU(2)\times
SU(2)$ FS model \cite{gibfreed}. Although we have not seen how to interpret this
solution in terms of intersecting branes its existence follows from the
1/4 supersymmetric $adS_3\times \bE^1$ `axiovac'. To see this,
one writes the $adS_3\times \bE^1$  metric in the  form (\ref{eq:threef}) and
reduces to D=3 in the $y$ direction. This yields a D=3 electrovac which can be
lifted to the D=4 electrovac with $adS_2\times\bE^2$ metric. Thus these 
two 1/4 supersymmetric solutions of the $SU(2)\times SU(2)$ FS model are dual to
each other. 
 
There are also other gauged supergravities. Many have now been provided with a
KK interpretation and, given such an interpretation, it is often possible to
interpret the KK compactification as the near-horizon geometry of a p-brane or,
as shown here, of intersecting branes. An exception is the gauged D=7
supergravity with topological mass term \cite{luca}. This theory has an adS
vacuum but no known KK interpretation, although it is tempting to suppose that
it is obtainable by some modification of the $S^3$ compactification.
Another outstanding exception is the D=6 $SU(2)$ gauged supergravity of Romans
\cite{romans}. This theory has an $adS$ vacuum with the exceptional
supergroup $F(4)$ as its isometry supergroup, but it has no known KK
interpretation. It is natural to suspect that it arises as the effective theory
in some compactification of D=10 supergravity. If so one might suppose that
it is again the near-horizon geometry of some intersecting brane solution, but 
no obvious candidate presents itself. We should also point out that there are
non-compact gaugings of D=4 N=8 supergravity that arise from `non-compact'
compactifications of D=11 supergravity \cite{hullw}, but the latter are not
known to occur as near-horizon geometries of any brane, or intersecting
brane, solutions. 

Consideration of the near-horizon geometries of branes and their intersections
has led to a number of compactifications of D=10 and D=11 supergravity theories
that were unknown in the heyday of Kaluza-Klein theory. The $S^3$ and
$S^3\times S^3$ compactifications of D=10 supergravity to domain walls
and electrovacs are examples. Another example is the $S^7$ compactification
of IIA supergravity to an $adS_3$ `linear dilaton' vacuum \cite{dgt}. 
The $S^3\times S^3$ compactification to the D=4 `axiovac' discussed
here similarly establishes a new 1/4 supersymmetric $S^3\times S^3\times S^1$
compactification of D=10 N=1 supergravity to D=3; the effective D=3 field
theory is presumably a matter-coupled adS supergravity. One may wonder whether
there are any new gauged supergravity theories that might be found in this way.
For example, the fact that the near horizon geometry of the linear intersection
of an M2-brane with an M5-brane is $adS_3 \times \bE^5 \times S^3$
\cite{kallosh,skenderis} means that there is an $S^3$ compactification of D=11
supergravity to D=8 preserving at least 1/4 supersymmetry.

\vskip 0.5cm
\noindent
{\bf Acknowledgements}: 
PKT gratefully acknowledges the support of the Iberdrola 
{\it Profesor Visitante} program. PMC thanks the members of the Faculty of
Physics at the University of Barcelona for hospitality. We are also grateful to
A. Chamseddine, D.Z. Freedman and C.M. Hull for helpful conversations
and to Robert Myers for pointing out an error in an earlier version of
the paper..

\end{document}